\documentstyle[epsf,graphicx]{mn}

\def\x2{$\chi^{2}$}

\def\ginga{{\it Ginga}~}
\def\asca{{\it ASCA}~}
\def\rosat{{\it ROSAT}~}

\def\x2{$\chi^{2}$}

\newbox\grsign \setbox\grsign=\hbox{$>$} \newdimen\grdimen \grdimen=\ht\grsign
\newbox\simlessbox \newbox\simgreatbox \newbox\simpropbox
\setbox\simgreatbox=\hbox{\raise.5ex\hbox{$>$}\llap
     {\lower.5ex\hbox{$\sim$}}}\ht1=\grdimen\dp1=0pt
\setbox\simlessbox=\hbox{\raise.5ex\hbox{$<$}\llap
     {\lower.5ex\hbox{$\sim$}}}\ht2=\grdimen\dp2=0pt
\setbox\simpropbox=\hbox{\raise.5ex\hbox{$\propto$}\llap
     {\lower.5ex\hbox{$\sim$}}}\ht2=\grdimen\dp2=0pt

\def\cunits{$\rm cm^{-2}$}
\def\funits{$\rm erg~cm^{-2}~s^{-1}$}

\begin{document}

\title[{\sl RXTE} observations of Markarian 3]
{\bf  {\sl RXTE} monitoring observations of Markarian 3}
\author[I. Georgantopoulos et al.]
{\Large I. Georgantopoulos$^1$,  I. Papadakis $^2$, 
 R.S. Warwick$^3$, D.A. Smith$^{4,5}$  G.C. Stewart$^3$ and 
R.G. Griffiths$^3$ \\
$^1$ National Observatory of Athens, Lofos Koufou, Palaia Penteli, 
15236, Athens, Greece \\
$^2$ Physics Department, University of Crete, Heraklion, Greece \\
$^3$ Department of Physics and Astronomy, University of Leicester, 
Leicester, LE1 7RH \\
$^4$ Laboratory for High Energy Astrophysics, Code 662, NASA/GSFC, 
Greenbelt, Maryland, MD20771, USA  \\
$^5$ Astronomy Department, University of Maryland, College Park, MD 20742, USA}

\maketitle
\begin{abstract}
We present Rossi X-ray Timing Explorer, monitoring observations of the Seyfert
2 galaxy Markarian 3 spanning a 200 day period during which time the 
source flux varied by a factor $\sim 2$ in the 4--20 keV bandpass. In broad 
agreement with earlier Ginga results, the average spectrum can be represented
in terms of a simple spectral model consisting of a very hard power-law 
continuum ($\Gamma \approx 1.1$) modified below $\sim 6$ keV by a high 
absorbing 
column ($N_H\sim 6\times 10^{23}$ \cunits) together with a high equivalent 
width Fe-K emission feature at 6.4 keV. The abnormally flat spectral index 
is probably the signature of a strong reflection component and we consider two
models incorporating such emission. In the first the reflected signal suffers 
the same absorption as the intrinsic continuum, whereas in the second the 
reflection is treated as an unabsorbed spectral component. In the former case,
we require a very strong reflection signal ($R~_{\sim}^{<}~3$) in order to
match the data; in addition variability of {\it both} the intrinsic power-law 
and the reflection component is required.
The unabsorbed reflection model requires a somewhat higher line-of-sight
column density to the nuclear source ($\sim 10^{24}$ \cunits), but in this 
case the reflected signal remains constant whilst the level of the intrinsic 
continuum varies. The latter description is consistent with the reflection 
originating from the illuminated far inner wall of a molecular torus, the 
nearside of which screens our direct view of the central continuum source.
\end{abstract}

\newpage

\section{INTRODUCTION}

In recent years the X-ray properties of Seyfert galaxies have been 
extensively studied (for a review see Mushotzky, Done \& Pounds 1993). 
It was soon realised that Seyfert 2 galaxies exhibit much lower
X-ray luminosities, at least in the soft X-ray band below $\sim 3$ keV,
than those typical of Seyfert 1 galaxies, a result which can now be 
explained in terms of the standard AGN unification model (Antonucci \& Miller 
1985). According to this paradigm, the nucleus (supermassive black hole, 
accretion disc and broad line region) has  basically the same structure 
in both types of object but, depending on the circumstances, can be hidden 
from viewed by a thick molecular torus (Krolik \& Begelman 1986). 
Specifically if the source is observed at a sufficiently high 
inclination angle,  and thus the line of sight 
intersects the torus, it would be classified as a Seyfert 
2, whereas for all other orientations it would be deemed to be a Seyfert 1. 
As well as obscuring the nucleus in the optical, the molecular torus can strongly
suppress soft X-ray emission through the process of photoelectric absorption
in cool atomic and molecular gas. In the limit of very high column 
densities, Thomson scattering will also diminish the more penetrating 
hard X-ray emission.

The X-ray spectra of Seyfert 2 galaxies as observed by \ginga, \asca and 
recently {\it BeppoSAX}~~have proved to be very complex (e.g. Awaki et al. 
1991; Smith \& Done 1996; Turner et al. 1997a; Griffiths et al. 1998). 
In broad terms most Seyfert 2 X-ray spectra can be well fitted by a power-law 
continuum (typically $\Gamma \sim 1.8$), plus an Fe-K emission line at 6.4 keV 
and a reflection component (e.g. Lightman 
\& White 1988, George \& Fabian 1991). This latter component, which
may be produced at the surface of the putative molecular torus,
flattens the observed continuum and can dominate the spectrum 
above $\sim 10$ keV.  In most Seyfert 2s the above emission components 
are viewed through a large absorbing column density,  typically  
$N_H>10^{23}$ \cunits. In some sources additional emission in
the form of a soft X-ray excess is observed below $\sim 3$ keV
probably as a result of scattering of the intrinsic power-law continuum by a
strongly photoionised medium. In order to observe such soft X-ray emission
it is clearly a requirement that the scattering medium should extend, 
in projection on the sky, well beyond the bounds of the obscuration 
of the molecular torus. 

In contrast to the recent progress in understanding the X-ray spectral 
characteristics of Seyfert 2 galaxies, our knowledge of their X-ray
variability properties remains very limited. According to the standard 
unification scenario, the hard X-ray continuum should vary with large 
amplitude in a similar way to that observed in Seyfert 1 galaxies 
(Mushotzky, Done \& Pounds 1993). However, in type 2 objects the
accretion disk, if present, is probably viewed at an acute angle and
also soft X-rays emanating directly from the nucleus are suppressed. 
Since the remaining reprocessed spectral components, namely the 
Fe-K line, the reflection signal and the soft excess, are likely to originate 
from regions of parsecs scale-size, it follows that significantly less 
variability might be expected in Seyfert 2 objects, at least in those parts of 
the spectrum where the reprocessing makes a substantial contribution to the 
overall flux.

In this paper, we focus on the properties of the Seyfert 2 galaxy Markarian 
3 (hereafter Mrk 3) which, at a redshift $z = 0.0137$, is one of the 
brightest and consequently most well-studied, members of its class. \ginga 
observations (Awaki et al. 1990; Awaki et al. 1991; Smith \& Done 1996) 
first revealed an abnormally flat power-law continuum emerging through a 
high obscuring column ($N_H\sim 6\times 10^{23}$ \cunits). A strong Fe line 
was also detected with a high equivalent width ($\approx 1.3$ keV). 
Mrk 3 has the hardest spectrum of all 16 Seyfert 2 studied by Smith \& Done 
(1996), significantly harder than the spectrum of Seyfert 1 galaxies, thus
challenging the standard unification models if the observed continuum actually 
corresponds to the underlying power-law in this source. However, several other
Seyfert 2 galaxies have been found to possess flat spectra and strong Fe lines 
(e.g. Reynolds et al. 1994; Maiolino et al. 1998; Iwasawa \& Comastri 1998) 
 with 
spectra generally indicating very heavy obscuration along the line of sight 
and also the presence of a strong Compton reflection component. 

Observations of Mrk 3 with the high spectral resolution afforded by the \asca 
SIS have resolved the Fe-K line into at least two components 
(Iwasawa et al. 1994). The dominant component at 6.4 keV has 
an equivalent width of 0.9 keV and a FWHM of $\sim 10^4$ $\rm km~s^{-1}$, 
while the second component at 7 keV has an equivalent width of 0.2 keV and 
appears to be narrower than the first. The same \asca observations 
(Iwasawa et al. 1994) require a spectral index of $\Gamma \approx 1.8$ 
but unfortunately the limited spectral bandpass (0.6--10 keV) of \asca 
provides only weak constraints on the properties of the intrinsic  
continuum. A re-analysis of the Mrk 3 spectrum using non-simultaneous 
\ginga, \rosat and \asca observations (Griffiths et al. 1998), covering a 
wide spectral band (0.1--30 keV), yielded a near canonical value for 
the power-law, $\Gamma \approx 1.7$, when either an additional
absorption edge at 8 keV (perhaps originating in a warm absorber), 
or reflection was included in the spectral model. 
 Recent observations with  {\it BeppoSAX}  (Cappi et al. 1999), 
 which extend the spectral coverage to 150 keV,  indeed 
 confirm the presence of a steep ($\Gamma\sim 1.8$) intrinsic power-law.    
Turner et al. (1997) have also re-analysed 
the \asca data and propose an alternative model in which the intrinsic
continuum is viewed through a very large absorbing column ($N_H>10^{24}$ 
\cunits) while the reflection component is unobscured (in contrast to the 
standard reflection scenario in which the direct power-law and reflection 
components are observed through the same $N_H$).
 The {\it BeppoSAX} observations (Cappi et al. 1999) 
 again support the above picture.  Such a model would be 
applicable, for example, if we have a direct, unobscured  view of the 
illuminated (far) inner walls of the torus. 

Time variability studies  can provide additional constraints on the geometry 
of the nucleus and the surrounding region. Comparison of the \ginga, 
{\it BBXRT} and the \asca measurements shows a decrease in the 2--10 keV
continuum flux by almost a factor of 3 (Iwasawa et al. 1994).
During the same period the Fe line flux decreased by a factor of 1.8 
(Griffiths et al. 1998),  substantially less than the continnum variation. 
However, Turner et al. (1997) examined the short-term variability using \asca 
observations but found no significant ($<$90 per cent) variability on 
timescales as short as one day. 

Here, we present the results of a X-ray monitoring campaign carried out on 
Mrk 3 by the Rossi X-ray Timing Explorer ({\sl RXTE}) mission, spanning a 
period of $\sim 200$ days. Our objective is to use the variability 
exhibited in the 4--20 keV band to place constraints on the geometry of 
the Mrk 3 nucleus and any surrounding gaseous media. The extended energy range 
of the {\sl RXTE} detectors also 
provides the opportunity to explore further the spectral composition 
of the X-ray emission emanating from Mrk 3. 

\section{THE OBSERVATIONS}

Mrk 3 was observed with the {\sl RXTE} between 25 December 1996  
and 6 July 1997. 
In total 12 observations were obtained, with a duration of about 5 ksec each. 
We have both Proportional Counter Array (PCA) and High Energy X-ray Timing 
Experiment (HEXTE) data but here we present the PCA analysis only. 
The PCA consists of five collimated (1$^{\circ}$ FWHM) Xenon proportional 
counter units (PCU). The PCU are sensitive to energies between 2 and 60 keV.
 However,  the effective area drops very rapidly below  4 and above 20 keV.  
The energy resolution  is 18 per cent at 6.6 keV 
 (Glasser, Odell \& Seufert 1994). 
The collecting area of each PCU is 1300 $\rm cm^2$. We extract PCU light 
curves and spectra from only the top Xenon layer in order to maximize the 
signal-to-noise ratio. We use only 
3 PCUs (0 to 2); the data from the other two PCU were discarded as these 
detectors were turned off on some occasions. 
The data were selected using standard screening criteria: we
exclude data taken at an Earth elevation angle of less than 10$^{\circ}$,
 pointing offset less than 0.01$^{\circ}$
 and during and 30 minutes after the satellite passage 
 through the South Atlantic Anomaly (SAA). 
 The  resulting total integration time is 59 ksec.
 In both the spectral and the timing analysis, we 
 use only data between 4 and 20 keV where the effective area 
 is the highest. 

We use the {\small PCABACKEST v2} routine of {\small FTOOLS v 4.1.1}
to generate the background models which take into account both the 
cosmic and the internal background. The internal background is estimated 
by matching the conditions of the observations with those in various model 
files. Most of the internal background is correlated with the L7 rate, the 
sum of 7 rates from pairs of adjacent anode wires. However, there is a 
residual background component correlated with recent passages 
from the SAA. Therefore, the use of a second, activation
component is also necessary. The level of the residual internal background
errors after background subtraction with  {\small PCABACKEST} is 
 about 20 per cent of the cosmic X-ray background $1\sigma$ fluctuations in the 
2-10 keV band.  The observation date for each dataset  
together with the observed background-subtracted 
count rate in the full 2-60 keV PCA energy band are given in Table 1.   

\begin{table}
\begin{center}
\caption{Log of the 12 {\sl RXTE} observations}
 \begin{tabular}{ccc}
Obs.No. & Date & Count Rate \\
        &      & $\rm ct~s^{-1}$ \\
\hline 
1 & 25/12/96  & 2.5  \\
2 & 17/02/97  & 3.2 \\
3 &  16/03/97 & 3.5 \\
4 & 21/03/97  & 3.8  \\
5 & 31/03/97  & 4.3  \\
6 & 04/04/97  & 4.0  \\
7 & 14/04/97  & 5.2  \\
8 & 15/04/97  & 5.0  \\
9 & 16/04/97  & 4.3  \\
10 & 17/04/97 & 5.0  \\
11 & 30/05/97 & 2.5  \\
12 & 06/07/97 & 2.3  \\ \hline 
\end{tabular}
\end{center}
\end{table}

\section{Time Variability}

\begin{figure*}
\rotatebox{0}{\includegraphics[height=13.0cm]{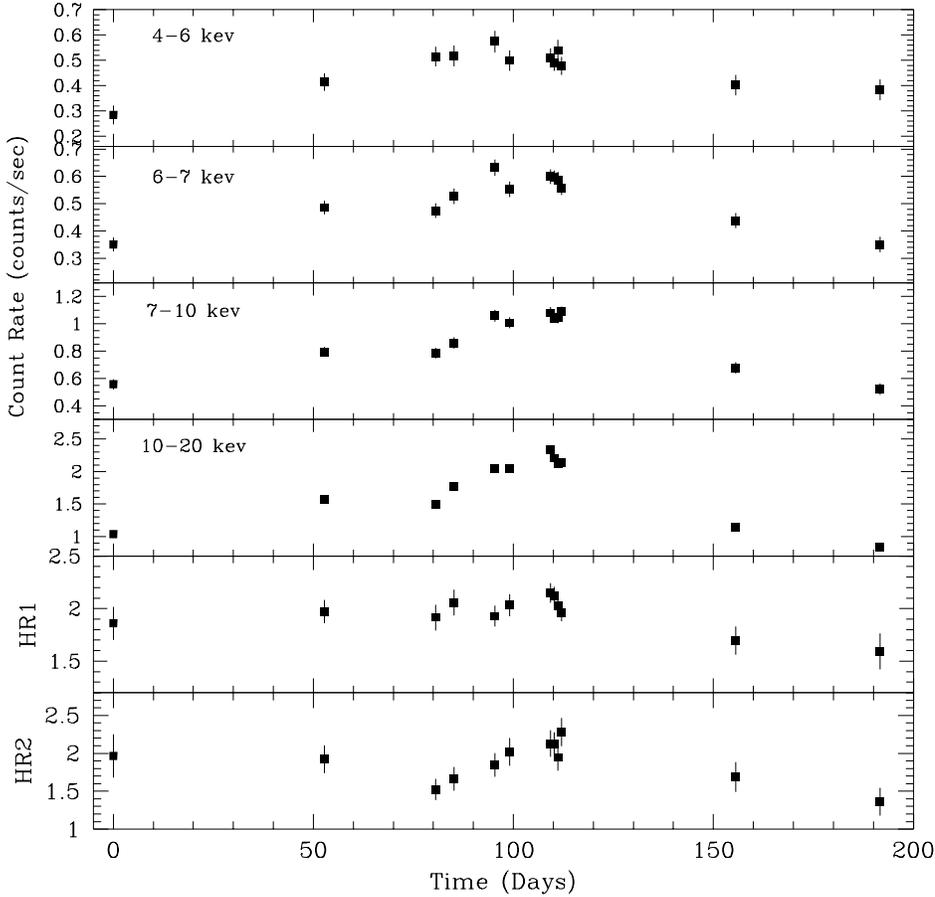}}
\caption{The background subtracted light curves in four different energy bands
(upper panels). Plots of the 10--20 keV/7--10 keV (HR1) and
 7--10 keV/4--6 keV (HR2) hardness ratios as a function of time are also 
shown (lower two panels).}
\label{lc}
\end{figure*}

Here, we address the issue of flux and spectral variability in a model 
independent way, using the background subtracted light curves. 
We divide the 4 to 20 keV range into 4 bands, namely 4--6 keV and 
7--10 keV, where the underlying power-law probably dominates the flux, 
6--7 keV where a significant fraction of the flux should originate  
from the Fe line at 6.4 keV  and finally 10--20 keV where the flux is 
quite possibly dominated by a reflection bump. The light curves for these  
bands are shown in Fig. \ref{lc} with $\pm1\sigma$ error bars. 
It is evident that there is variability in all four
bands, with a minimum-to-maximum amplitude of at least a factor of two. 
This result is statistically significant at a high 
level of confidence; a constant value for the count rate gives $\chi^2$ values 
of  33, 134, 307 and 850 (11 degrees of freedom) for the 4--6, 6--7, 7--10 and 
10--20 keV bands respectively. The rms variation in the respective 
light curves, after correcting for the variance  due to the photon statistics,
is 15, 17, 22 and 28 per cent.  The apparent increase in the fractional 
variability amplitude towards higher energy might arise, for example, if a 
less variable or constant spectral component contributes preferentially
to the softer bands;
however spectral analysis does not substantiate this simple picture (see the
next section).

We investigated the possibility of spectral variability by plotting
the ratios of the  10--20 keV/7--10 keV  and 7--10 keV/4--6 keV count rates, 
denoted as HR1 and HR2 respectively, as a function of time (see Fig. \ref{lc}).
Both the HR1 and HR2 ratios provide some evidence for such temporal 
variability; against the constant ratio hypothesis the $\chi^2$ values 
are respectively 20 and 27  (for 11 degrees of freedom), corresponding to 
only a 5 and 0.5 per cent chance probabilities. Hence, the data
do suggest the presence of subtle spectral variations, but unfortunately
do not provide any strong pointers to a preferred spectral model.

\section{Spectral Analysis}

The PCA data from each observation were binned to give a minimum
of 20 counts per channel. All data below 4 and above 20 keV were ignored
due to their poor signal-to-noise ratio.  By discarding the data 
below 4 keV we also avoid the complications associated with the soft X-ray 
excess in this source (e.g. Griffiths et al. 1998). 
The spectral fitting 
analysis was carried out using the {\small XSPEC v.10} software package
on the basis of ``joint simultaneous fits'' to the 12 {\sl RXTE} 
observations.

\begin{figure*}
\rotatebox{270}{\includegraphics[height=12.5cm]{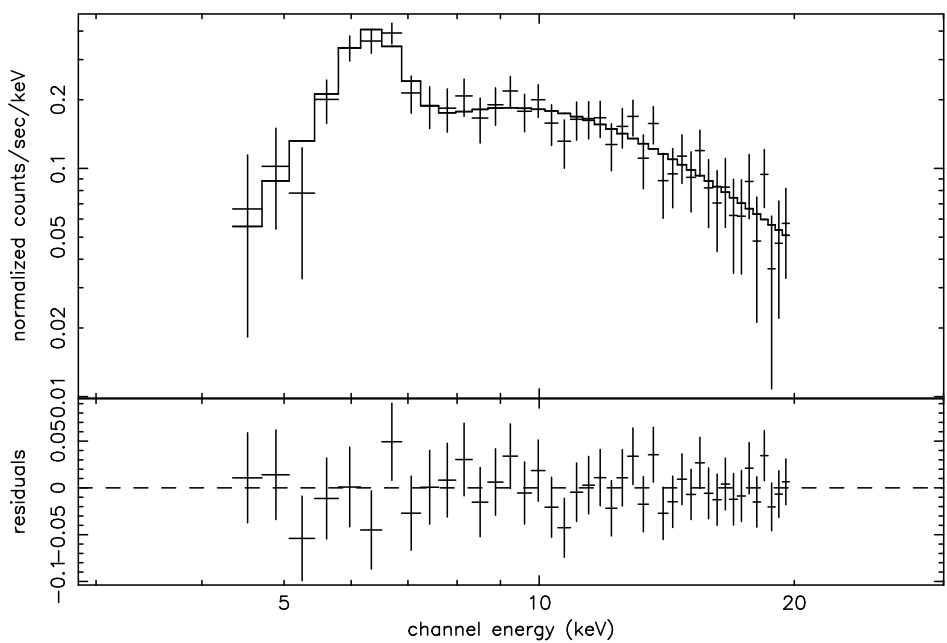}}
\caption{The {\sl RXTE} PCA spectrum of Mrk 3 from observation 1.
The solid line corresponds to the best fitting version of model A  after
folding through the instrument response. The lower panel shows the 
corresponding residuals to the fit.}
\label{typical}
\end{figure*} 
 
Following previous \ginga and \asca results, we first employ a very simple
spectral model consisting of a power-law continuum, with photon spectral 
index $\Gamma$, modified by absorption in a column density, $N_H$, of
cool neutral material. A Gaussian line was also included 
to account for Fe-K emission.  
For simplicity the line energy and intrinsic width were fixed at
the values obtained from \asca by Griffiths et al. (1998), 
{\it i.e.} $E_{line} = 6.38$ keV and $\sigma_{line}=0.1$ keV (since consistent 
values were obtained in free fits of these parameters).
  The values of the $N_H$, the photon index and the normalization of the 
 Fe line are free parameters but they are tied to the same value. However, 
 the normalizations of the power law are allowed to vary freely. 
The results of fitting this spectral model are presented in Table 2 as model A,
where the errors correspond to the 90 per cent confidence level for one
interesting parameter. The derived photon index is very flat ($\Gamma\approx 
1.1$), consistent with the original Smith \& Done (1996) 
analysis of the \ginga data. The resulting $\chi^2$ of 365 for 489 degrees of 
freedom (dof) implies 
an excellent fit (although the errors arising from the background model, 
may have been somewhat overestimated).
Fig. \ref{typical} shows a typical PCA count rate spectrum from a single
observation, together with the best-fitting model prediction
and the resulting residuals.  
 As can be seen from Fig. \ref{typical} we clearly observe signal up to 20 keV,
 even in the case of the lowest flux observations.  

The derived Fe-K line flux was $4.7\pm0.84\times 10^{-5} 
\rm~photon~s^{-1}~cm^{-2}$ implying
an  equivalent width varying from 1.4 keV (observation 12) to 0.4 keV 
(observation 2). The flux of the Fe line is entirely consistent 
with that obtained by Iwasawa et al. (1994). 
When the normalisation of the Fe-K component was ``untied''
across the set of observations, the resulting  $\chi^2$ reduced to 357 for 
478 dof, but this is not a statistically significant 
improvement (according to the F-test for 11 additional parameters).
We conclude that there is no strong evidence for variability in the Fe-K line 
flux in the {\sl RXTE} data. In fact this conclusion holds for all
spectral models (models A-D) considered in this section, and so in each case 
we have kept the Fe-K normalisation tied to a single value for all the 
datasets. 

\begin{table*}
\begin{center}
\caption{Results from the spectral fitting of the 12 {\it RXTE} datasets}
\begin{tabular}{ccccccc}
Model  & $N_H$ & $\Gamma$  & R & $E_{\small edge}$ &
 $\tau$ &  $\chi^2$/dof  \\
\hline 
{\small A} & $63^{+4}_{-4}$ & $1.06^{+0.04}_{-0.06}$ 
        &  - & - & - & 365/489  \\
{\small B} & $75^{+3}_{-3}$ & 1.8 & $0.1^{+1.0}_{-0.1}-3.4^{+0.5}_{-0.5}$ 
 & - & - & 337/478 \\
{\small C} &
$110^{+6}_{-6}$ & $1.85^{+0.09}_{-0.09}$ & 
 $0.7^{+0.14}_{-0.14}-2.5^{+0.5}_{-0.5}$ & - & - & 322/489  \\ 
{\small D} & $74^{+3}_{-3}$ & $1.3^{+0.14}_{-0.14}$ & - & 8.1 & 0.24 & 
376/489  \\
\hline
\end{tabular}
\end{center}
\end{table*}
  
The next step was to investigate whether there is any evidence for 
changes in either the photon spectral index or the column density. 
The result of spectral fitting with each of these parameters in turn 
``untied'' was a $\chi^2$ of 337 and 335 respectively 
for 478 dof. In both cases the change in $\chi^2$  (compared to model A) is 
statistically significant at over 99 per cent confidence. The range of 
the apparent variation in the 
spectral index is $\Gamma = 0.94^{+0.16}_{-0.16}$ to $1.54^{+0.35}_{-0.29}$ 
 while the best  fit column density is $N_H=66^{+5}_{-9}\times 10^{22}$ $\rm cm^{-2}$. 
 Similarly, when the $N_H$ is untied we obtain  
$N_H= 47^{+6}_{-6}$ to $ 69^{+5}_{-5} \times
10^{22} \rm~cm^{-2}$ while the best fit 
 photon index is $\Gamma= 1.14 ^{+0.11}_{-0.11}$.  
 This confirms the evidence from the hardness ratios
 for underlying spectral variability. Note that there is clearly a dependence 
 of the photon index on the column density in the sense that 
 the data cannot easily discriminate between a flat photon index 
 and a high column density. 

\begin{figure} 
\rotatebox{270}{\includegraphics[height=10.0cm]{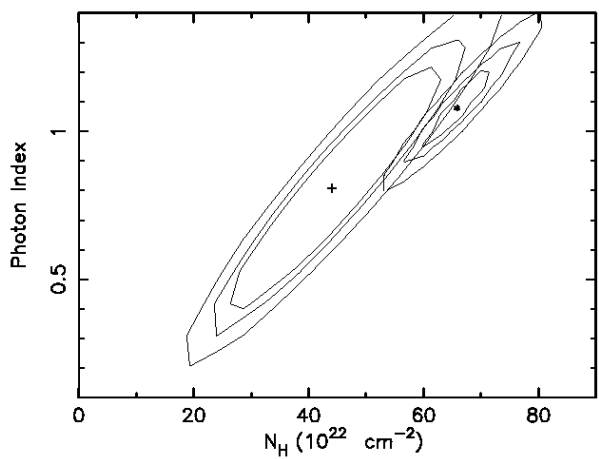}}
\caption{The $\Gamma-N_H$ contours for the high-flux (rightmost)
and the low-flux data (leftmost); in each case the three levels correspond 
to 68, 90 and 99 per cent confidence contours.} 
\label{contours}
\end{figure}

As an additional test,  we checked for possible spectral variations
correlated with the X-ray brightness of the source. For this purpose
we separated the observations into high and low-states on the basis a flux 
threshold of $3\times 10^{-11}$ \funits~in the 4--20 keV band. Observations 
4--10 were thereby classified as high state, with observations 1--3 and 
11--12 comprising the low-state. The results (based on the model A
prescription) are summarised in Fig. \ref{contours} where we plot the 
joint 68, 90 and 99 per cent confidence contours in the $\Gamma$ versus $N_H$
plane for the high-flux and the low-flux states.  
We see that although the best fit centroids are offset, the 90 per cent 
confidence contours show considerable overlap, suggesting 
that we cannot be confident that either the continuum slope or column density
shows any consistent change with increasing flux.

\begin{figure} 
\rotatebox{0}{\includegraphics[height=8.0cm]{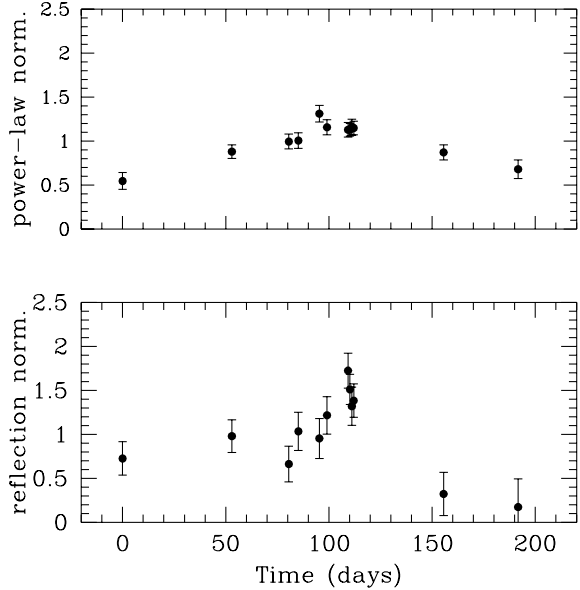}}
\caption{The normalization of the power-law (top) and the 
reflection components (bottom)  in the case of the 
standard reflection model (model B). The errors 
correspond to a 90 per cent confidence level.}
\label{refl}
\end{figure}

Although, the model A prescription defined above gives an 
acceptable fit in terms of the $\chi^{2}$ statistic, there is evidence from 
both \asca and \ginga observations (Griffiths et al. 1998) 
 and recently {\it BeppoSAX} observations (Cappi et al. 1999) that the 
anomalously flat power-slope derived for Mrk 3 is due to the presence of
Compton reflection. The next step in the current analysis was therefore 
to include a reflection component in the spectral modelling. Specifically 
we use the {\small PEXRAV} model (Magdziarz \& Zdziarski 1995) in 
{\small XSPEC}.
We initially assume that both the reflection component and the power-law are 
absorbed by the cold gas column density according to the standard reflection 
prescription. The strength of the reflection component is governed by the 
parameter R, representing the strength of the reflected signal 
relative to the level of the incident power-law continuum. Following the 
 results of Griffiths et al. (1998),  
 we fix the inclination angle for the disk at 
$i=60^{\circ}$; this large inclination angle implies that a 
 large part of the reflection component originates in the torus. 
 We also set the spectral index of the power-law continuum to 
($\Gamma=1.8$) following the results of Griffiths et al. (1998) and 
  Cappi et al. (1999). Note that 
 without this constraint the fit reverts to a very 
flat power-law slope and negligible reflection. Our initial approach
was to tie the {\it effective normalisation} of the reflection signal to a 
single value across the set of observations. The resulting best-fitting 
model gave a $\chi^{2}$ of 395 for 489 dof and required values of 
$R$ varying from 0.7$\pm 0.2$--2.1$\pm0.9$. 
 In this model even though the reflection component 
has a fixed level,  R changes since the normalisation of the
intrinsic power-law varies from observation to observation. However, 
when we allow the normalisation of the reflection component to vary freely, 
we obtain a $\Delta\chi^2\approx 58$ which is highly statistically 
significant. Details of this fit are summarised in Table 2 (Model B) and 
Fig. \ref{refl} shows the derived temporal variation in the normalisation 
of both the power-law and reflection  components. 
There is clearly a suggestion that the latter responds to variations in the 
former with any lag between the two being $ _{\sim}^{<} 1$ month. 
Unfortunately there is insufficient data to set a more precise constraint.

As noted earlier an alternative reflection model for Mrk 3 was proposed by 
Turner et al. (1997) in which the intrinsic power-law is seen through an 
increased column but the reflection is largely unobscured. The application of
such a model to the {\sl RXTE} data gives the results presented in Table 2 
(model C) when  the reflection normalisations are tied to a single value.
The best-fitting power-law slope is 
$\Gamma=1.85^{+0.09}_{-0.09}$, consistent with the canonical AGN power-law, 
and $N_H =1.1 \times 10^{24}$ \cunits, {\it i.e.} approximately 30\%
higher than the value obtained earlier (as model B).
 Note that at such high column densities the effects of Thomson 
 scattering become important. Then the derived column density from fitting 
 simple absorption models will overestimate the true column (Leahy et al. 1989). 
 The model above gives the lowest 
reduced $\chi^2$ of all the models we consider ($\chi^2= 322/489$). 
Comparison with the standard reflection model (in the 
case where the reflection component has a constant normalisation) 
using the F-test, suggests that the Turner et al. model 
represents a better fit to the data at the 90 per cent 
confidence level.  Interestingly in this second reflection scenario 
we obtain only a small improvement in the $\chi^2$ ($\Delta \chi^2\sim 6$)
when the normalisation of the reflection component is allowed to vary
across the observations. Thus in this description the reflection
component essentially remains constant.

Finally, we note that Griffiths et al. (1998) suggest an alternative 
explanation for the abnormally flat spectrum of Mrk 3. These authors include
an additional absorption edge near 8 keV in their spectral model which
serves to steepen the slope determined for the underlying power-law component.
The edge feature could originate in a putative warm absorber 
(e.g. Reynolds 1997; George et al. 1998) which in the case of Mrk 3
may also produce the scattered soft excess flux observed below 3 keV
(assuming an extensive distribution of this hot medium in the nuclear region
of the galaxy).  We have therefore also considered the effect of including
such a feature in the spectral fitting of the {\it RXTE} datasets. 
The results are given in Table 2 (model D),  where we have fixed the edge 
energy and optical depth at the values obtained by Griffiths et al. 
({\it i.e.} $E_{edge}  =8.1$ keV, $\tau = 0.24$). In this case the
power-law remains flat ($\Gamma \approx 1.3$) and the best fit is actually 
worse than that obtained in the absorbed power-law model (model A).
We conclude that the absorption edge alone is not sufficient to explain the 
apparently anomalously hard spectrum measured for Mrk 3 by {\sl RXTE}.


\section{DISCUSSION}

The {\sl RXTE} observations presented in this paper confirm the finding
of previous studies namely that the X-ray continuum emanating from
Mrk 3 is exceptionally hard, at least within the 4-20 keV bandpass.
The preferred interpretation of this flat spectrum is that this source
exhibits particularly strong Compton reflection and we find that
two variants on this reflection theme are broadly consistent with the 
{\sl RXTE} data both in terms of the average spectrum and the 
observation to observation spectral variability.

In the standard reflection description (e.g. Griffiths et 
al. 1998) the intrinsic power-law continuum, the reflection component and 
the Fe-K emission are all affected by photoelectric absorption in
a large column density of cool absorbing gas ({with 
$N_H \sim 7\times 10^{23}$ \cunits). In this model both the continuum
and the reflected component vary together; with any lag in the response
of the latter constrained to  $ _{\sim}^{<} 1$ month. 
In contrast, the {\sl RXTE} data show no evidence 
for variability in the Fe line flux.  
 It is quite plausible that the (bulk of the) Fe-K flux
and the reflection signal originate in different regions. For example
a significant fraction of the Fe-K line flux 
 might originate in a very extended region which is optically
thin to Fe-K photons whereas the reflection could arise in a partially
covering screen of optically thick clouds situated within a light month of
the nucleus.

In the alternative version of the reflection model (e.g. Turner et al. 1998)
our line of sight to the reflecting material is largely unobscured, although
the nucleus itself is covered by a very substantial screen of absorption
($N_H\sim 10^{24}$ \cunits). This model actually gave the best fit to 
the average spectrum  and a fairly canonical value for the intrinsic 
power-law slope. Also the only temporal variation required is in the level of 
the underlying continuum. This leads to arguably the most plausible
explanation of the anomalously hard spectrum observed in Mrk 3, namely
that we see strong Compton reflection from the far illuminated wall of 
a putative molecular torus in its nucleus. Presumably our line of sight to 
this region passes over the nearside of the torus without intercepting
anything like the column density that lies directly in front of the 
central nuclear source.  

Future monitoring observations by satellites such as XMM 
will greatly improve the photon statistics as well as the spectral resolution 
and thus are expected to shed further light on the detailed geometry of the central region
of Mrk 3. Such observations will in fact provide a critical and detailed
test of current unification schemes. 

\section{Acknowledgements}

We thank the anonymous referee for many useful comments and 
suggestions. RGG acknowledges support from PPARC in the form 
of a research studentship.

\section*{References}

Antonucci, R.R.J., Miller, J.S., 1985, ApJ, 297, 621 \\
Awaki, H., Koyama, K., Kunieda H., Tawara, Y., 1990, Nature, 346, 544 \\
Awaki, H., Koyama, K., Inoue, H., Halpern, J.P., 1991, PASJ, 43, 195 \\ 
Cappi, M., et al., 1999, A\&A, in press \\
George, I.M., Fabian, A.C., 1991, MNRAS, 249, 352 \\
George, I.M., Turner, T.J., Netzer, N., Mandra, K., Mushotzky, R.F., 
 Yaqoob, T., 1998, ApJS, 114, 73 \\
Glasser, C.A., Odell, C.E., Seufert, S.E., 1994, 
 IEEE Trans. Nucl. Sci., 41, 4\\
Griffiths, R.E., Warwick, R.S., Georgantopoulos, I., Done, C., Smith, D.A., 
1998, MNRAS, 298, 1159 \\
Iwasawa, K., Yaqoob, T., Awaki, H., Ogasaka, Y., 1994, PASJ, 46, L167 \\ 
Iwasawa, K., Comastri, A., 1998, MNRAS, 297, 1219 \\
Krolik, J.H., Begelman, M.C., 1986, ApJ, 308, 55 \\
 Lee, J.C., Fabian, A.C., Reynolds, C.S.,  Iwasawa, K.,  Brandt, W.N., 
 1998, MNRAS, 300, 583 \\
Leahy, D.A., Matsuoka, M., Kawai, N., Makino, F., 1989, MNRAS, 236, 603 \\
Lightman, A.P., White, T.R., 1988, ApJ, 335, 57 \\
Maiolino, R.,  Salvati, M., Bassani, L., Dadina, M.,  Della Ceca, R.,
  Matt, G., Risaliti, G., Zamorani, G., 1998, A\&A, 338, 781 \\ 
Magdziarz, P., Zdziarski, A., 1995, MNRAS, 273, 837 \\
Mushotzky, R.F., Done, R.F., Pounds, K.A., 1993, ARA\&A, 31, 717 \\
Smith, D.A., Done, C., 1996, MNRAS, 280, 355 \\
Nandra, K., Pounds, K.A., 1994, MNRAS, 268, 405 \\
Nandra, K., Mushotzky, R.F., Yaqoob, T., George, I.M., Turner, T.J., 1997, 
MNRAS, 284, L10 \\
Reynolds, C.S., 1997, MNRAS, 286, 513 \\
Reynolds, C.S., Fabian, A.C., Makishima, K., Fukazawa, Y., Tamura, T., 
1994, MNRAS, 268, L55 \\
Turner, T.J., George, I.M., Nandra, K., Mushotzky, R.F., 1997,
ApJS, 113, 23 \\
Turner, T.J., George, I.M., Nandra, K., Mushotzky, R.F., 1997,
ApJ, 488, 164 \\

\end{document}